\documentstyle[12pt,epsf]{article}
\newcommand{\beq}{\begin{equation}}
\newcommand{\eeq}{\end{equation}}

\newcommand{\sla}[1]{#1 \!\!\! /}
\newcommand{\nn}{\nonumber}
\newcommand{\ul}[1]{\underline{#1}}
\newcommand{\qs}{\hat{q}^2}
\newcommand{\rt}{\rho_\tau}
\newcommand{\rj}{\rho_c}
\newcommand{\te}{\Theta_\nu}
\newcommand{\sq}{\sqrt{y^2-4\rt}}

\newcommand{\matel}[3]{\langle #1|#2|#3\rangle}

\newcommand{\ra}{\rightarrow}

\renewcommand{\Im}{\mbox {Im}\:}

\def\lsim{\mathrel{\rlap{\lower3pt\hbox{\hskip0pt$\sim$}}
    \raise1pt\hbox{$<$}}}         
\def\gsim{\mathrel{\rlap{\lower4pt\hbox{\hskip1pt$\sim$}}
    \raise1pt\hbox{$>$}}}         

\newcommand{\ga}{\gamma}
\newcommand{\de}{\delta}

\newcommand{\GeV}{\,\mbox{GeV}}

\begin{document}
\begin{titlepage}
\renewcommand{\thefootnote}{\fnsymbol{footnote}}
\begin{flushright} \bf
PITHA 95/2\\
arch-ive/9502207 \\
February 95
\end{flushright}
\vspace*{.4cm}
\begin{center} \LARGE
{\bf Tau Polarization in $ \Lambda_b \ra X_c \tau \bar{\nu}$ and $B \ra X_c
\tau \bar{\nu}$}
\end{center}
\vspace*{.3cm}
\begin{center} \Large{M. Gremm\footnote{e-mail address: {\it
gremm@acds15.physik.rwth-aachen.de}},
 G. K\"opp, L. M. Sehgal\footnote{ e-mail address: {\it
sehgal@acphyz.hep.rwth-aachen.de}} }\\
\end{center}
\vspace*{.4cm}
\begin{center}
{\bf III. Physikalisches Institut (A)}\\
{\bf RWTH Aachen, D-52074 Aachen, Germany}
\end{center}
\vspace{.3cm}
\begin{center}
{\Large{\bf Abstract}}
\end{center}
\vspace*{.2cm}

We discuss the longitudinal and transverse $\tau$-polarization in inclusive
decays
of hadrons containing $b$-quarks. The calculation is performed by means of an
OPE
in HQET. Some mathematical difficulties in
calculating transverse polarizations are explained. Numerical results are
presented for longitudinal and for transverse polarizations, both in and
perpendicular to the decay plane.

\end{titlepage}
\addtocounter{footnote}{-2}

\section{Introduction}

In this paper we present a detailed discussion of lepton polarization in the
inclusive
decay of the $B$-meson and  the $\Lambda_b$. Our calculations make use of
the heavy quark effective theory (HQET) and an operator product
expansion (OPE). These techniques were initially developed  for
massless leptons in the final state \cite{georgi} \cite{bigi0} \cite{wise}
\cite{blok}.
Recently massive leptons in $B$ decays have also been considered \cite{kor}
\cite{korn} and their longitudinal polarization has been calculated in
\cite{falk}.
We will supplement this discussion by a calculation of the
longitudinal $\tau$-polarization in the decay $\Lambda_b \ra X_c \tau
\bar{\nu}$
and the transverse $\tau$-polarization in $B$ and $\Lambda_b$ decays.

In the calculation of transverse polarizations one encounters certain divergent
integrals
if the decay width is analysed as a function of the momentum transfer $q^2$.
In \cite{lig} the same divergences have been encountered
in the calculation of a CP violating matrix element for transverse $\tau$-
polarization. In that work the authors introduce a regularization procedure
in order to obtain finite results. We will show that these divergences
can be bypassed by using an alternative parametrization
that yields mathematically well defined quantities. A similar mathematical
problem
is responsible for the occurrence of $\de$-functions  and their derivatives in
the differential
decay widths calculated in \cite{wise}. Our parametrization removes this
problem as well.
Thus all seemingly unphysical aspects of the inclusive rates, i.e.
$\de$-functions
and divergences in polarizations, are of mathematical origin and can be
avoided.

The transverse polarization can be viewed
as an additional test of HQET and the operator product expansion. For
instance, a CP conserving matrix element for $B$-meson decays should not
produce any polarization
transverse to the decay plane. This is confirmed by our calculations.
On the other hand, for polarized $\Lambda_b$ decay, there is a nontrivial
transverse $\tau$-polarization which is correlated with the $\Lambda_b$-spin,
and contains components both in and perpendicular to the decay plane.
Experimentally polarizations can be measured and compared to theoretical
predictions without any knowledge of the CKM matrix element $|V_{cb}|$. Also
the fifth power of the somewhat hazy $b$-quark mass does not appear in the
results, thereby
reducing the uncertainties of the predictions further.
On a more technical level
the calculation of transverse polarization is an interesting test of a theory
that
claims to describe small corrections to the free quark model.

Mainly to establish notation, we briefly review some of the basic steps
of this calculation. If not stated otherwise we will use the notations
established in \cite{wise}. The decay widths are of the form
\beq
d\Gamma=2|V_{cb}|^2G_F^22\pi
L_{\mu\nu}H^{\mu\nu}\frac{d^3P_\nu}{(2\pi)^32P_\nu^0}
	\frac{d^3P_\tau}{(2\pi)^32P_\tau^0}
\eeq
Subsequently we will use them normalized to the free quark model total decay
width $\Gamma_0=\frac{|V_{cb}|^2G_F^2m_b^5}{192\pi^3}$. The hadron tensor is
given by the imaginary part of the matrix element of a transition operator.
\beq
\label{tramp}
H_{\mu\nu}=-\frac{1}{\pi}\Im T_{\mu\nu}
\eeq
Both sides of this equation can be decomposed into form factors multiplied
by Lorentz structures. The most general decomposition excluding terms violating
time reversal invariance is
\begin{eqnarray}
T^{\mu\nu}&=&-g^{\mu\nu}T_1+v^\mu v^\nu T_2-
	i\epsilon^{\mu\nu\alpha\beta}v_\alpha q_\beta T_3
	+q^\mu q^\nu T_4+(q^\mu v^\nu+q^\nu v^\mu)T_5 \\
&&-(qs_h)\left( -g^{\mu\nu}S_1+v^\mu v^\nu S_2-
	i\epsilon^{\mu\nu\alpha\beta}v_\alpha q_\beta S_3
	+q^\mu q^\nu S_4+(q^\mu v^\nu+q^\nu v^\mu)S_5 \right) \nn \\
&&+(s_h^\mu v^\nu+s_h^\nu v^\mu)S_6
 +(s_h^\mu q^\nu+s_h^\nu q^\mu)S_7
 +i\epsilon^{\mu\nu\alpha\beta}v_\alpha s_{h\beta}S_8
 +i\epsilon^{\mu\nu\alpha\beta}q_\alpha s_{h\beta}S_9 \nn
\end{eqnarray}
$q$ is the total momentum of the lepton pair, $v$ is the 4-velocity of the
decaying hadron and $s_h$ is the hadron spin vector.
All form factors are functions of $q^2$ and $qv$. For the
hadrons under consideration here they can be calculated by means of an
operator product expansion in the heavy quark
effective theory. $L_{\mu\nu}$ is the usual lepton tensor
\beq
\label{lptu}
L^{\mu\nu}= 8 \left(P_\tau^\mu P_\nu^\nu+P_\tau^\nu P_\nu^\mu-
	g^{\mu\nu}(P_\tau P_\nu) +
	i\epsilon^{\mu\nu\alpha\beta}P_{\tau\alpha}P_{\nu\beta}
	\right)
\eeq
For polarized leptons it is
convenient to introduce
\begin{eqnarray}
L_{\mu\nu}^s&=&Tr[(\sla{P}_\tau+m_\tau) \frac{1+\ga^5 \sla{s}_\tau }{2} \ga^\mu
(1-\ga^5)\sla{P}_\nu \ga^\nu (1-\ga^5)] \\
\label{lptens}
L_{\mu\nu}^p&=&\frac{1}{2}(L_{\mu\nu}^s-L_{\mu\nu}^{-s}) \nonumber \\
 &=&-4m_\tau \left( P_\nu^\nu s_\tau^\mu + P_\nu^\mu s_\tau^\nu -
g^{\mu\nu} (P_\nu s_\tau)-i\epsilon^{\mu\nu\alpha\beta}P_{\nu\alpha}
s_{\tau\beta}
\right)
\end{eqnarray}
and to calculate the matrix element obtained by contracting the hadron tensor
with (\ref{lptens}).
Then the $\tau$-polarization in $s_\tau$ direction is given by
\beq
2\frac{d\Gamma^p}{d\Gamma}
\eeq
where $d\Gamma^p$ is the decay width calculated with the lepton tensor
(\ref{lptens}), and $d\Gamma$ with (\ref{lptu}).

The paper is organized as follows: In section II the OPE is discussed. Section
III is devoted to the calculations of the longitudinal $\tau$-polarization
in the decay $\Lambda_b \ra X_c \tau \bar{\nu}$. A detailed discussion of the
transverse polarization in $\Lambda_b$ decays, including a discussion of the
mathematical difficulties, will be given in section IV. The same calculations
are repeated for $B$-meson decays in section V. Section VI contains the
numerical evaluation.

\section{The Operator Product Expansion}

We recapitulate briefly the operator product expansion in the context of HQET
\cite{georgi} \cite{wise} \cite{blok} \cite{korn} \cite{bigi}.
A convenient way to perform the expansion makes use of the fact that the
heavy quark propagator can be written as a geometric series.
The starting point for constructing an OPE is the transition operator
\beq
 \hat{T}^{\mu\nu} = -i\int d^4 x {\rm e}^{-iqx} T[ j^{\mu\dagger} (x)
j^\nu (0)]
\eeq
which takes the form
\beq
\label{teq}
 \hat{T}^{\mu\nu}= \bar{b} \ga^{\mu} P_L
\frac{1}{P_b\!\!\!\!/-q\!\!\!/-m_c}\ga^{\nu}P_Lb
\eeq
for $b \ra c$ transitions. Here $P_L$ is the left handed projection operator,
$P_b$ is the $b$-quark momentum and $q=P_\tau+P_\nu$ is the total momentum of
the
$\tau$ and the $\nu$. To obtain the transition amplitude $T_{\mu\nu}$ in
(\ref{tramp}) this
operator is placed between $B$-meson or $\Lambda_b$ states.
The OPE expresses the propagator in the transition
operator through a sum of operators. To construct this expansion one interprets
the $b$-quark momentum as an operator $P_b^\mu \ra i\partial^\mu+g_s A^\mu$
which
contains the gluon field $A$ because the propagating $c$-quark cannot be
viewed as free. It is
exposed to the background field generated by interactions with
the light degrees of freedom in
the hadron. The full QCD $b$-field in (\ref{teq}) is related to the field in
HQET through
\beq
b = {\rm e}^{-im_bvx}(1+\frac{iD\!\!\!/}{2m_b}+\cdots)b_v
\eeq
The $b$-quark momentum has been decomposed into a large part due to the
hadron's motion and a small part due to interactions between the heavy quark
and the light quarks: $P_b=m_bv+k$. $v$ is the 4-velocity of the meson and $k$
is the momentum due to interactions between the quarks. Between $b$-quark
fields the
$b$-momentum operator $\hat{P_b}$ can be replaced: $\hat{P_b} \ra
m_bv+\hat{k}$.
Now the gluon field is contained in $\hat{k} = i\partial+g_sA$.
The operator $\hat{k}$ produces only the
residual momentum of ${\cal O}(\Lambda_{QCD})$. It obeys the commutator
relation:
$[\hat{k}^\mu,\hat{k}^\nu]=ig_sG^{\mu\nu}$, which implies that the gluon tensor
is of second order in $k$. The OPE can be performed if the propagator is
rewritten as \\
\beq
\label{prop}
 \frac{1}{m_b\sla{v}+\sla{\hat{k}}-\sla{q}-m_c} =
 (\sla{V}_0+\sla{\hat{k}}+m_c) \frac{1}{(V_0+\hat{k})^2+
\frac{1}{2}g_s\sigma^{\alpha\beta}G_{\alpha\beta}-m_c^2}
\eeq
with $V_0=m_bv-q$ and $\sigma^{\alpha\beta}=\frac{i}{2}(
\ga^\alpha\ga^\beta-\ga^\beta\ga^\alpha)$.
Defining $\Delta_0
= V_0^2-m_c^2$ the fraction can be rewritten in a form suitable for
expansion in a geometric series
\begin{eqnarray}
\frac{1}{(V_0+\hat{k})^2+
\frac{1}{2}g_s\sigma^{\alpha\beta}G_{\alpha\beta}-m_c^2}
&=& \frac{1}{\Delta_0} \frac{1}{1+\frac{2V_0\hat{k}+\hat{k}^2+
\frac{1}{2}g_s\sigma^{\alpha\beta}G_{\alpha\beta}}{\Delta_0}} \nn \\
&=& \frac{1}{\Delta_0} \sum_{n=0}^\infty \left(-\frac{2V_0\hat{k}+\hat{k}^2+
\frac{1}{2}\sigma^{\alpha\beta}G_{\alpha\beta}}{\Delta_0}\right)^n
\end{eqnarray}
This expansion corresponds to an expansion in $\frac{\hat{k}}{m_b}$ since
$\Delta_0$ is of ${\cal O} (m_b^2)$. Inserting this expansion in (\ref{prop})
and using the algebra of the $\ga$-matrices gives the expansion of the
propagator
in powers of $\hat{k}$. Expressions containing the charm mass do not contribute
because of the projection operators in the amplitude (\ref{teq}).
\begin{eqnarray}
\frac{1}{\sla{V_0}+\sla{\hat{k}}-m_c}& =&
\frac{\sla{V_0}}{\Delta_0}+\frac{\sla{\hat{k}}}
{\Delta_0}-2\frac{\sla{V_0}V_0\hat{k}}{\Delta_0^2}-
2\frac{\ga^\delta V_0^\tau\hat{k}_{(\delta}\hat{k}_{\tau)}}{\Delta_0^2}
\nonumber \\
&&-\frac{\sla{V_0}\hat{k}^2}{\Delta_0^2}+
4\frac{\sla{V_0}(V_0\hat{k})^2}{\Delta_0^3}
+\frac{g_s}{2\Delta_0^2}V_{0\alpha} \ga_\beta \ga_5
\epsilon^{\alpha\beta\delta\tau}G_{\delta\tau}+ {\cal O}(\hat{k}^3)
\end{eqnarray}
The brackets around indices indicate a symmetrized expression.
These operators are identical to those found in \cite{wise}.
The method of
obtaining the matrix elements of these operators is explained in detail in
that paper. We will not repeat that procedure here.

\section{Longitudinal $\tau$-Polarization in $\Lambda_b$ decays}

The calculation of the longitudinal $\tau$-polarization in decays of
polarized $\Lambda_b$'s
 can be done in precisely the same way as the inclusive rates \cite{wise}
\cite{blok} and
the longitudinal polarization for $B$-meson decays \cite{falk} were calculated.
We will use that method here without describing the details of the procedure.
For massive leptons a number of form factors in addition to those given
in \cite{wise} are necessary to calculate the decay rates. In the notation
of Manohar and Wise \cite{wise}, which we use throughout this paper, they are
(to
${\cal O}(\hat{k}^2)$)

\begin{eqnarray}
T_4&=&\frac{4}{3}m_b(K+G)\frac{1}{\Delta_0^2} \\
T_5&=&-\frac{1}{2}\frac{1}{\Delta_0} -
\frac{1}{3}m_bK(4m_b+5qv)\frac{1}{\Delta_0^2} -
\frac{5}{3}m_bGqv\frac{1}{\Delta_0^2} \nn \\
& &+\;\;\;\;\;\frac{4}{3}m_b^2K(q^2-(qv)^2)\frac{1}{\Delta_0^3} \\
S_4&=&0 \\
S_5&=&-\frac{2}{3}m_bK\frac{1}{\Delta_0^2} \\
S_7&=&\frac{1}{2}(1+\epsilon)\frac{1}{\Delta_0}+
m_bK(qv+\frac{2}{3}m_b)\frac{1}{\Delta_0^2}-
\frac{4}{3}m_b^2K(q^2-(qv)^2)\frac{1}{\Delta_0^3}
\end{eqnarray}
$\epsilon$ is a spin symmetry breaking parameter. Its value can be estimated
through sum rule calculations. Sum rules  \cite{bigi} \cite{wise}
also give a value for the constant
\beq
K=-\matel{H_b}{ \bar{b_v}\frac{(iD)^2}{2m_b^2} b_v }{H_b}=0.01
\eeq
We use this value for $B$-mesons and $\Lambda_b$ even though it is by no
means a universal constant.
The matrix element of the chromomagnetic operator between $B$-meson states is
\beq
G=\matel{B}{\bar{b_v}\frac{g_sG_{\alpha\beta}\sigma^{\alpha\beta}}{4m_b^2}
	b_v}{B}=-0.0065
\eeq
but vanishes between $\Lambda_b$ states.
For the longitudinal polarization the spin projection vector $s_\tau$ in
(\ref{lptens}) is set to
\beq
s_\tau=\left(\frac{|\underline{P}_\tau |}{m_\tau},\frac{P_\tau^0
\underline{P}_\tau}{m_\tau |\underline{P}_\tau|} \right)
\eeq
The decay widths, normalized to the free quark total decay width and
transformed
to normalized variables $x=\frac{2P_\nu^0}{m_b}, y=\frac{2P_\tau^0}{m_b},
\qs=\frac{q^2}{m_b^2}$
\begin{eqnarray}
& & \frac{1}{\Gamma_0}d\Gamma=
  \frac{3}{2}\frac{1}{m_b}L_{\mu\nu}H^{\mu\nu}dxdyd\hat{q}^2d\cos(\Theta_s) \\
& &\frac{1}{\Gamma_0} d\Gamma^p=\frac{3}{2}\frac{1}{m_b}
	L_{\mu\nu}^p H^{\mu\nu}dxdyd\hat{q}^2d\cos(\Theta_s)
\end{eqnarray}
can be calculated as shown in \cite{wise}.
$\Theta_s$ is the angle
between the spin vector of the $\Lambda_b$ and the direction of the $\tau$
3-momentum. After integration over $x$ and $\hat{q}^2$ the spin-independent
part of the decay width can be written symbolically as
\beq
\frac{1}{\Gamma_0}\frac{d\Gamma}{dyd\cos(\Theta_s)}=
	2\pi\left(A_0+A_1 K+\left(B_0(1+\epsilon)+B_1 K\right)\cos(\Theta_s)
	\right)
\eeq
The functions $A_0, A_1, B_0$ and $B_1$ are given by
\begin{eqnarray}
\label{A0}
A_0&=& \frac{1}{2\pi}\sqrt{y^2-4\rt} \left( (-3y^2+6y(1+\rt)-12\rt)x_0^2
	\right. \nn \\
&& \left. + (y^2-3y(1+\rt)+8\rt)x_0^3 \right) \\
\label{A1}
A_1&=& \frac{1}{2\pi}\frac{\sq}{(1+\rt-y)^2}
	\left( -2(y-2)(y-2\rt)(y^2-4\rt) \right. \\
& & \left. -4(y^4-3y^3(1+\rt)+y^2(5+6\rt+5\rt^2)-12(1+\rt)\rt y \right. \nn \\
& & \left. -8\rt(1-4\rt+\rt^2))x_0 \right. \nn \\
& & \left. +2(2y^4-8(1+\rt)y^3+(15+28\rt+15\rt^2)y^2-52\rt(1+\rt)y \right. \nn
\\
& & \left. -18\rt(1-6\rt+\rt^2))x_0^2
\right. \nn \\
& & \left. -\frac{4}{3}(y^4-5y^3(1+\rt)+2y^2(5+11\rt+5\rt^2)-40\rt(1+\rt)y
	\right. \nn \\
& & \left.- 2\rt(5-38\rt+5\rt^2))x_0^3 \right) \nn \\
\label{B0}
B_0&=&  \frac{1}{2\pi} \left(-3(y^3-2y^2\rt-4y\rt+8\rt^2)x_0^2 \right. \\
&&  \left. +(y^3+y^2(1-3\rt)-4y\rt-4(1-3\rt)\rt)x_0^3\right)  \nn \\
\label{B1}
B_1&=& \frac{1}{2\pi}\frac{y^2-4\rt}{(1+\rt-y)^2}
	 \left( -2(y-2\rt)(y^2-4\rt)  \right. \\
& & \left. -4(y^3-(4+3\rt)y^2+(5\rt+11)\rt y
	+6\rt(1-3\rt))x_0 \right. \nn \\
& & \left. +2(2y^3-2(3+4\rt)y^2+(-5+22\rt+15\rt^2)y+24\rt(1-2\rt))x_0^2
	\right. \nn \\
& & \left. -\frac{4}{3} (y^3-(2+5\rt)y^2+(-5+11\rt+10\rt^2)y+6\rt(3-5\rt))x_0^3
	\right) \nn
\end{eqnarray}
The polarization-dependent width has a similar decomposition
\beq
\frac{1}{\Gamma_0} \frac{d\Gamma^p}{dy d\cos(\Theta_s)}=
	2\pi\left(a_0+a_1K+\left(b_0(1+\epsilon)+b_1 K\right)\cos(\Theta_s)
		\right)
\eeq
with
\begin{eqnarray}
a_0&=& \frac{1}{2\pi}(y^2-4\rt) \left(
\frac{3}{2}(y-2)x_0^2+\frac{1}{2}(3-\rt-y)x_0^3\right) \\
a_1&=& \frac{1}{2\pi}\frac{(y^2-4\rt)}{(1+\rt-y)^2} \left( (y^3-2y^2-4\rt
y+8\rt) \right. \\
& & \left. +2(y^3-y^2(3+4\rt)+(5+11\rt) y+6\rt(\rt-3))x_0 \right. \nn \\
& & \left. +(-2y^3+2y^2(4+3\rt)-y(15+22\rt-5\rt^2)-24\rt(\rt-2))x_0^2 \right.
\nn \\
& & \left. +\frac{2}{3}(y^3-y^2(5+2\rt)+y(10+11\rt-5\rt^2)-6\rt(5-3\rt))x_0^3
\right)
	\nn \\
b_0&=&  \frac{1}{2\pi}\sq \left(\frac{3}{2}(y^2-4\rt)x_0^2-\frac{1}{2}(y^2
	+y(1+\rt)-8\rt)x_0^3 \right)  \\
b_1&=&  \frac{1}{2\pi}\frac{\sq}{(1+\rt-y)^2} \left(
	 (y^2-4\rt)^2+2(y^2-4\rt)(y^2-4y(1+\rt)+12\rt)x_0 \right. \\
& & \left. - (2y^4-6y^3(1+\rt)+y^2(-5+8\rt-5\rt^2)+44\rt y(1+\rt)
	\right. \nn \\
&& \left. +\rt(10-124\rt+10\rt^2))x_0^2
	\right. \nn \\
& & \left. +\frac{2}{3}(y^4-2y^3(1+\rt)-5y^2(1+\rt^2)+28\rt y(1+\rt)
	+2\rt(5-38\rt+5\rt^2))x_0^3 \right)  \nn
\end{eqnarray}
In all of these expressions
$x_0=1-\frac{\rho_c}{1+\rt-y},
\rho_c=\frac{m_c^2}{m_b^2},\rt=\frac{m_\tau^2}{m_b^2}$.\\
We have checked that our result for $d\Gamma$ reproduces the result in
\cite{wise} for vanishing lepton mass. The part proportional to $K$ in
the polarized decay width reproduces the result for $B$ decays in \cite{falk}.

\section{Transverse Polarization in $\Lambda_b$ decays}

As already indicated one encounters certain technical difficulties in
the calculation of transverse polarizations
if the parametrization in terms of the momentum transfer $\qs$ is used. To
illustrate the origin of these difficulties we will discuss the transverse
$\tau$-polarization for the decays of polarized $\Lambda_b$ in some detail.

\begin{figure}[t]
\begin{center}
\mbox{\epsfxsize 6cm \epsfysize 4cm \epsffile{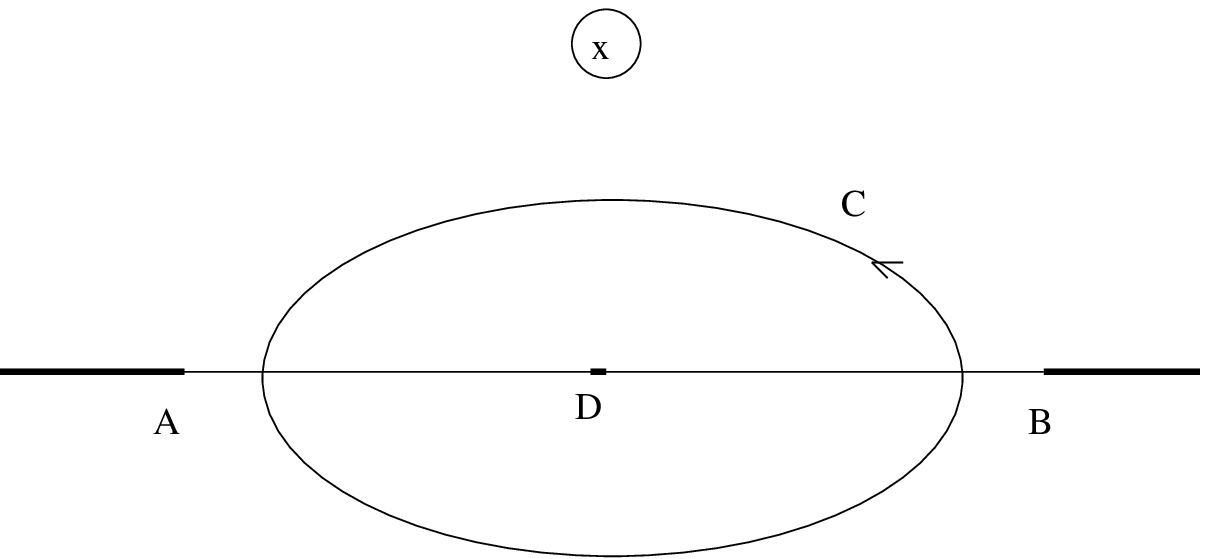} }
\end{center}
\caption{. }
\label{branch}
\end{figure}

We choose the rest frame of the hadron as our coordinate system and the
direction of the
$\tau$ emission as $z$-axis. In this system the vectors take the form
\begin{eqnarray}
P_\tau= \left( \begin{array}{c} P_\tau^0 \\ 0 \\ 0 \\ |\ul{P}_\tau| \end{array}
	\right),\;\;\;\;\;
P_\nu= \left( \begin{array}{c} P_\nu^0 \\
	|\ul{P}_\nu|\cos(\phi_\nu)\sin(\Theta_\nu) \\
	|\ul{P}_\nu| \sin(\phi_\nu)\sin(\Theta_\nu) \\
	|\ul{P}_\nu|\cos(\Theta_\nu) \end{array}
	\right)
\nn \\
s_h= \left( \begin{array}{c} 0 \\
	\cos(\phi_\nu+\psi)\sin(\Theta_s) \\
	\sin(\phi_\nu+\psi)\sin(\Theta_s) \\
	\cos(\Theta_s) \end{array}
	\right),\;\;\;\;\;
s_\tau= \left( \begin{array}{c} 0 \\
	\cos(\phi_\nu+\de)\\
	\sin(\phi_\nu+\de)\\
	0 \end{array}
	\right)
\end{eqnarray}
$s_h$ is the spin vector of the hadron. $\psi\;(\de)$ is the azimuthal
angle between the hadron (lepton) spin and the neutrino momentum.
This parametrization of the
spin projection vector has the advantage that transverse polarizations
in arbitrary orientations relative to the decay plane can be determined.
$\de=0$
gives the transverse polarization in the decay plane, while $\de=\frac{\pi}{2}$
corresponds to $s_\tau$ perpendicular to that plane. The differential decay
width  in scaled variables is given by
\begin{eqnarray}
\label{dw}
\frac{1}{\Gamma_0}d\Gamma^p &=&
	\frac{3}{8m_b\pi}L^p_{\mu\nu}H^{\mu\nu}
  \de\left(\qs-\rt-\frac{1}{2}x(y-\sq \cos(\Theta_\nu))\right) \nn \\
 &&\cdot x\sq d\qs dx dy d\psi
	d\cos(\Theta_s)d\cos(\Theta_\nu)
\end{eqnarray}
A $\de$-function relating $\qs$ to $\cos(\Theta_\nu)$ has been included to
transform phase space and the matrix element to the same set of variables.
The matrix element $L_{\mu\nu}^pT^{\mu\nu}$ may be written symbolically as
\beq
L_{\mu\nu}^pT^{\mu\nu}=\left( f_1\frac{1}{\Delta_0}+f_2\frac{1}{\Delta_0^2}
	+f_3\frac{1}{\Delta_0^3} \right)
\eeq
with $f_i=f_i(x,y,\cos(\Theta_\nu),\sin(\Theta_\nu))$.
The functions $f_i, i=1,2,3$ are analytic in the complex $x$-plane.
Each power of $\Delta_0$ contributes a pole of corresponding order.
To obtain the decay width (\ref{dw}) as a function of $\qs$ instead of
$\cos(\Theta_\nu)$ one must integrate over $\Theta_\nu$.
This integration produces two Heaviside functions
\beq
\Theta \left( \frac{2(\qs-\rt)}{y-\sqrt{y^2-4\rt}}-x\right)
\Theta \left( x-\frac{2(\qs-\rt)}{y+\sqrt{y^2-4\rt}}\right)
\eeq
and the replacements
\begin{eqnarray}
\cos(\Theta_\nu) & \ra & -2\frac{\hat{q}^2-\rho_\tau - \frac{1}{2}xy}
	{x\sqrt{y^2-4\rho_\tau}}\\
\label{wurz}
\sin(\Theta_\nu) & \ra & 2\frac{\sqrt{( \hat{q}^2-\rho_\tau)xy-\rho_\tau
x^2-(\hat{q}^2-\rho_\tau)^2}}{x\sqrt{y^2-4\rho_\tau}}
\end{eqnarray}
In the scaled variables the poles have the form

\beq
\label{pol}
\frac{1}{(1-\rho_c-x-y+\hat{q}^2)^n},\; n=1,2,3
\eeq
The imaginary part of $L_{\mu\nu}T^{\mu\nu}$ and $L^p_{\mu\nu}T^{\mu\nu}$
can be taken by performing a
contour integration around the poles in the complex $x$-plane.
The apparent poles at $x=0$ are compensated by the factor
$|P_\nu|=\frac{1}{2}m_bx$
in the neutrino momentum $P_\nu$. Thus only (\ref{wurz})
alters the analytic structure of
the integrand by introducing two branch cuts in addition to the poles. The
analytic structure of the parts containing the square roots
 is sketched in Fig. \ref{branch}. $D$ is the position of the pole. The branch
cuts are chosen to extend from the points $A$ and $B$ to infinity. $C$
is the contour for the integration. As long as the pole and the ends of the
branch cuts are well separated the contour integration can be performed with
the residue theorem. This yields the
replacement of the poles by $\de$-functions as a means of
taking the imaginary part of the matrix element \cite{wise}.
A necessary condition for
this technique to be applicable is the analyticity of the function multiplying
the pole in the vicinity of the pole.

The subsequent integration over $\qs$ corresponds to moving the poles and the
ends of
the branch
cuts. For $\qs$ equal to its minimal or maximal value in the $\qs$ integration,
the pole and the end of the left or the right branch cut fall
together, i.e $A=D$ or $B=D$. For $\qs$ arbitrarily close to these values
the contour integral
is not defined because the contour is pinched between two non analytic
points. Correspondingly the integral over $\qs$ is well defined only if the
limits of the integral do not approach the maximum or minimum possible values.
If the limits are included the integral diverges if one replaces the poles with
$\de$-functions because the necessary condition for that replacement is no
longer fulfilled.

For poles of first order only the result stays finite for all values of $\qs$
because integrals over first order poles embedded in branch cuts can be
defined.
The calculation of transverse polarizations in the free quark model
produces only poles of first order as can be seen by setting $K$ and
$\epsilon$ zero in the form factors.
Therefore one does not encounter any divergences in that case.

The problem mentioned above
can be avoided by integrating (\ref{dw}) over $\qs$ instead of
$\cos(\Theta_\nu)$. Then the $\de$-function in (\ref{dw}) yields
\beq
\label{tfktn}
\Theta(x)\Theta(\sq)\Theta(1-\cos(\Theta_\nu))\Theta(\cos(\Theta_\nu)-1)
\eeq
The replacement
(\ref{wurz}) is never made, so no branch cuts are introduced. Setting
\beq
\qs = \rt+\frac{1}{2}x\left(y-\sqrt{y^2-4\rt}\cos(\Theta_\nu)\right)
\eeq
in the form factors leads to modified expressions for the poles.

\begin{eqnarray}
&&\frac{1}{(1-\rt-x-y+\qs)^n} \nn \\
&&\;\;\;\;\;\;\;\;\;\; = \frac{2^n}{(2-y+\sqrt{y^2-4\rt}\cos(\Theta_\nu))}
		\frac{1}{\left(2\frac{1-\rj+\rt-y}
		{2-y+\sqrt{y^2-4\rt}\cos(\Theta_\nu)}-x\right)^n}
\end{eqnarray}
For the maximal value of the $\tau$-energy $y_{max}=1+\rt-\rj$ the pole is
located at $x=0$. Strictly speaking a contour integral around the pole is
not defined in this case because $\Theta(x)F(x)$, $F(x)$ analytic,
does not possess an analytic
continuation for $x < 0$. However, since the derivatives of this function
are well defined in the sense of distributions, one can
treat this expression as analytic at the price of introducing its derivatives,
containing $\de$-functions, into the energy spectra. Substituting
\begin{eqnarray}
&& \frac{1}{\left(2\frac{1-\rj+\rt-y}
		{2-y+\sqrt{y^2-4\rt}\cos(\Theta_\nu)}-x\right)^n} \nn \\
&&\;\;\;\;\;\;\; \ra
\frac{(-1)^{n-1}}{(n-1)!}\de^{(n-1)}\left(2\frac{1-\rj+\rt-y}{2-y+
		\sqrt{y^2-4\rt}\cos(\Theta_\nu)}-x\right)
\end{eqnarray}
under the integral over $x$
to obtain the imaginary part directly leads to a matrix element that can be
integrated easily.
In this parametrization the expressions multiplying the Heaviside function
contain
a factor $x^2$ or a higher power of $x$ together with poles of at most third
order. Since for $x \ge 0$ $\Theta(x)x^n$ has the same finite
derivatives of up to $n$-th order as $x^n$, the Heaviside function can be
omitted in all expressions containing poles of up to order $n+1$
\footnote{
If it is not omitted it produces $\de$-functions with vanishing coefficients.
In the parametrization in terms of $\qs$ the factors $x^n, n \ge 2$, are
missing.
Therefore the $\de$-functions contribute in that case.}.
All expressions we consider in this paper fulfill this condition.
Thus after integration over $x$ no terms containing $\de$-functions survive.
To illustrate this a decay width differential in $y$ and $\cos(\te)$
is given in the appendix.
The integration over $x$ is trivial and the remaining integration can be
performed with a standard substitution.
It yields a decay rate of the structure

\begin{eqnarray}
\frac {1}{\Gamma_{0}} \frac{d\Gamma}{dy d\psi d\cos(\Theta_s)}&=&
	A_0+A_1K+\left(B_0(1+\epsilon)+B_1K\right)\cos(\Theta_s) \\
	&&+\left( C_0 (1+\epsilon)+C_1K\right)\sin(\Theta_s) \nn
\end{eqnarray}
for unpolarized leptons. The functions $A_0,A_1,B_0$ and $B_1$ are the same as
in equations (\ref{A0})(\ref{A1})(\ref{B0}) and (\ref{B1}).
\begin{eqnarray}
C_0&=& \frac{3}{8}\sqrt{1+\rt-y} \sqrt{y^2-4\rt}\cos(\psi)
		\left( (2y-4\rt)x_0^2 + (-y+2\rt)x_0^3\right) \\
C_1&=&\frac{\sqrt{y^2-4\rt}\cos(\psi)}{(1+\rt-y)^{3/2}}
 	\left( \frac{1}{2}\left(y^3-2\rt y^2-4\rt y+8\rt^2\right)  \right. \\
&& \left. +\frac{1}{4}(y^3-2(5+2\rt)y^2+4\rt(3\rt+8)y +8\rt(2-7\rt)) x_0
\right. \nn \\
&& \left.
		+\frac{3}{16} (-5y^3+2(10+9\rt)y^2-4\rt(16+7\rt)-24\rt(1-4\rt)) x_0^2
	\right. \nn \\
&& \left.
		+\frac{5}{32}(3y^3-2(5+6\rt)y^2+4\rt(9+4\rt)y \right.  \nn \\
&& \left. +8\rt(1-6\rt)) x_0^3
	\right) \nn
\end{eqnarray}
For polarized leptons the spin dependent part of the decay width reads
\begin{eqnarray}
\frac{1}{\Gamma_0}
	\frac {d\Gamma^p} {dy d\psi d\cos(\Theta_s)} &=&
	a_0+a_1K+\left(b_0(1+\epsilon)+b_1K\right)\cos(\Theta_s)  \\
&&	+\left(c_0(1+\epsilon)+c_1K\right)\sin(\Theta_s) \nn
\end{eqnarray}
\begin{eqnarray}
\label{a0}
a_0&=& \sqrt{\rt} \frac{3}{16} \sqrt{y^2-4\,\rt} \sqrt{1+\rt-y} \cos(\de) (y-2)
 x_0^3  \\
\label{a1}
a_1&=& \sqrt{\rt} \frac{1}{64}\frac{ \sqrt{y^2-4\rt}
\cos(\delta)}{(1+\rt-y)^{3/2}}
	\left( 24(y-2)(y^2-4\rt)  x_0  \right. \nn \\
& & \left.+4((1-15\rt)y^2 -2(1-29\rt)y+8((2-11\rt+2\rt^2))  x_0^2 \right. \nn
\\
& & \left.
	+5(-3y^3+2(6+5\rt)y^2-4(4+9\rt)y+8\rt(6-\rt)) x_0^3 \right) \\
b_0&=& \sqrt{\rt} \frac{3}{16}(y^2-4\rt) \sqrt{1+\rt-y} \cos(\de)x_0^3\\
b_1&=& \sqrt{\rt}\frac{(y^2-4\rt)\cos(\delta)}{64(1+\rt-y)^{3/2}}
	\left( 24 (y^2-4\rt)  x_0 \right. \\
&& 	\left. -12 ((1+5\rt)y+4(1-4\rt)) x_0^2
	\right. \nn  \\
& & 	\left. -5(3y^2-2(3+5\rt)y-4(1-6\rt)) x_0^3
	\right) \nn \\
c_0&=& -\frac{1}{\pi} \sqrt{\rt} \sqrt{y^2-4\rt}(1+\rt-y) \cos(\delta)
\cos(\psi) x_0^3  \\
c_1&=&
 \sqrt{\rt} \frac{\sq}{\pi (1+\rt-y)} \left(
	-2\cos(\delta) \cos(\psi)(y^2-4\rt)  x_0 \right. \\
& & \left.+  ( (\cos(\delta)\cos(\psi)-\cos(\de-\psi))y^2 \right. \nn \\
& & \left.+(-(1-3\rt)\cos(\delta)\cos(\psi)+(3+\rt)\cos(\delta-\psi))y
	\right. \nn \\
& & \left.
	+4(1-3\rt)\cos(\delta)\cos(\psi)-2(1+\rt)\cos(\delta-\psi) )  x_0^2
	\right. \nn \\
& & \left.
	+\frac{2}{3} \cos(\delta)\cos(\psi) (2y^2-5(1+\rt)y+12\rt) x_0^3
	\right) \nn
\end{eqnarray}

In \cite{lig} the same technical problem has been encountered in a somewhat
different context. We have checked that our method yields the same result
as the regularization procedure introduced in that paper.

\section{Transverse Polarization in $B$-meson decays}

The calculation of the transverse $\tau$-polarization in the decays of
$B$-mesons
proceeds along precisely the same lines. The matrix element is somewhat
simplified because of the absence of a hadron spin but the terms containing
the chromomagnetic operator can no longer be ignored.
Only the parts of $T^{\mu\nu}$ that do not contain the hadron spin contribute
to this decay.
The necessary equations,
definitions and techniques have been discussed in \cite{wise} \cite{blok}
\cite{falk}
or in this paper so
we will simply present our results for the polarized and unpolarized decay
width.
\beq
\label{Bg}
 \frac{1}{\Gamma_0} \frac{d\Gamma}{dy}= 4\pi\left(A_0+A_1K+A_2G\right)
\eeq
The functions $A_0$ and $A_1$ are identical to those in the $\Lambda_b$
decay rate into unpolarized leptons (\ref{A0})(\ref{A1}), while $A_2$ is
given by
\begin{eqnarray}
A_2&=&\frac{1}{\pi}\frac{\sq}{1+\rt-y} \left(
	\left( 5y^3-10(1+\rt)y^2+4(3+4\rt)y-16\rt(2-\rt)\right)x_0 \right. \\
&& \left. -\left( 5y^3-(17+15\rt)y^2+(24+46\rt)y-70\rt+18\rt^2\right)x_0^2
	\right. \nn \\
&& \left. +\frac{5}{3}\left(y^3-4(1+\rt)y^2+2(3+7\rt)y-4\rt(5-\rt)\right)x_0^3
	\right) \nn
\end{eqnarray}
The decay rate into transversely polarized leptons has the same decomposition
\beq
\label{Bgp}
\frac{1}{\Gamma_0} \frac{d\Gamma^p}{dy}= 4\pi(a_0+a_1K+a_2G)
\eeq
\begin{eqnarray}
a_2&=&\frac{1}{2}\sqrt{\rt}\frac{\sq\cos(\de)}{\sqrt{1+\rt-y}}
	\left( -\left( y-2 \right)x_0 \right. \\
&& \left.+\frac{1}{8}\left(-15y^2+50y-8(7-2\rt)\right)x_0^2 \right. \nn \\
&& \left. +\frac{5}{16}\left(5y^2-16y+4(4-\rt)\right)x_0^3
	\right) \nn
\end{eqnarray}
the functions $a_0, a_1$ being identical to (\ref{a0})
and (\ref{a1}).
The unpolarized decay width (\ref{Bg}) reproduces the result of \cite{kor}
\cite{korn}.
For the choice $\de=\frac{\pi}{2}$, i.e. $s_\tau$
perpendicular to the decay plane, the transverse polarization vanishes.

\section{Numerical Evaluation}

\begin{figure}[t]
\epsfxsize=352pt
\epsfysize=313pt
\begin{center}
\mbox{\epsffile{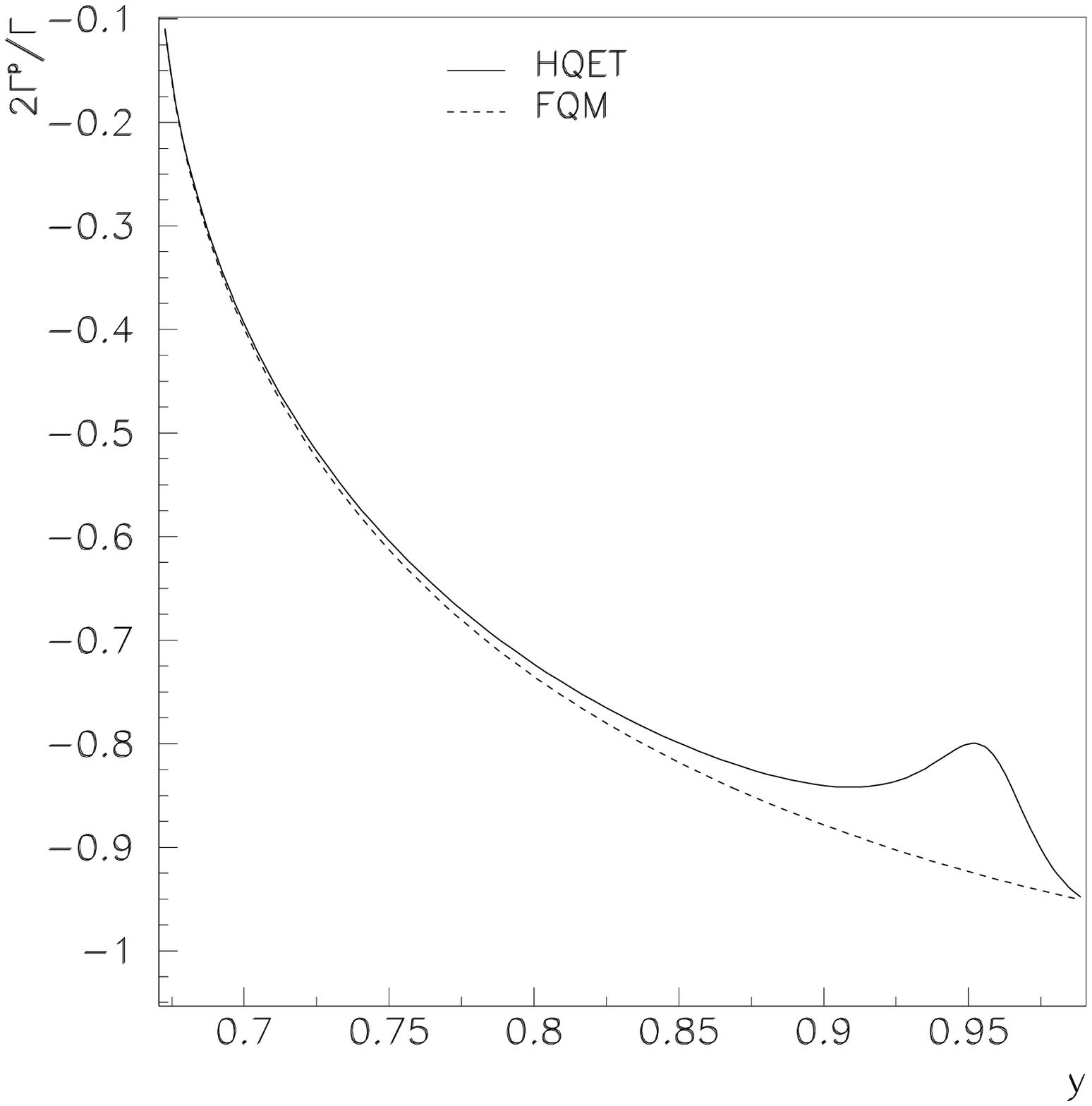}}
\end{center}
\caption{. }
\label{lpol}
\end{figure}

For the numerical evaluation of the spectra we have used the following
parameters: $m_b=5.3\GeV, m_c=1.85\GeV, m_\tau=1.777\GeV$. The HQET parameters
are set to $\epsilon=0, K=0.01$ and for meson decays $G=-0.0065$. Unlike the
energy spectra the polarizations are finite over the whole kinematically
allowed range of the $\tau$-energy. In Fig.~\ref{lpol} the longitudinal
$\tau$-polarization in polarized $\Lambda_b$ decays is shown for the
hadron spin perpendicular to the $\tau$-momentum. Other orientations of the
hadron spin yield qualitatively similar graphs. For
comparison the free quark model (FQM) prediction is included. The HQET
corrections are small. Only near the end of the spectrum they
become noticeable but in that region the OPE begins to break down.
This produces the endpoint divergences in the $\tau$-energy spectra which
have to be removed by applying smoothing functions before the results are
physically meaningful. We expect the peak in Fig. \ref{lpol} to be due to
the fact that we used the HQET spectra without smoothing them. If smoothed
spectra are used it probably will be less pronounced.
For the average polarization ${\cal P}=2\frac{\Gamma^p}{\Gamma}$ we obtain
\beq
{\cal P} = -0.72
\eeq
This value is in the same range as the longitudinal polarization in $B$-meson
decays calculated in \cite{falk}.

\begin{figure}[t]
\epsfxsize=352pt
\epsfysize=313pt
\begin{center}
\mbox{\epsffile{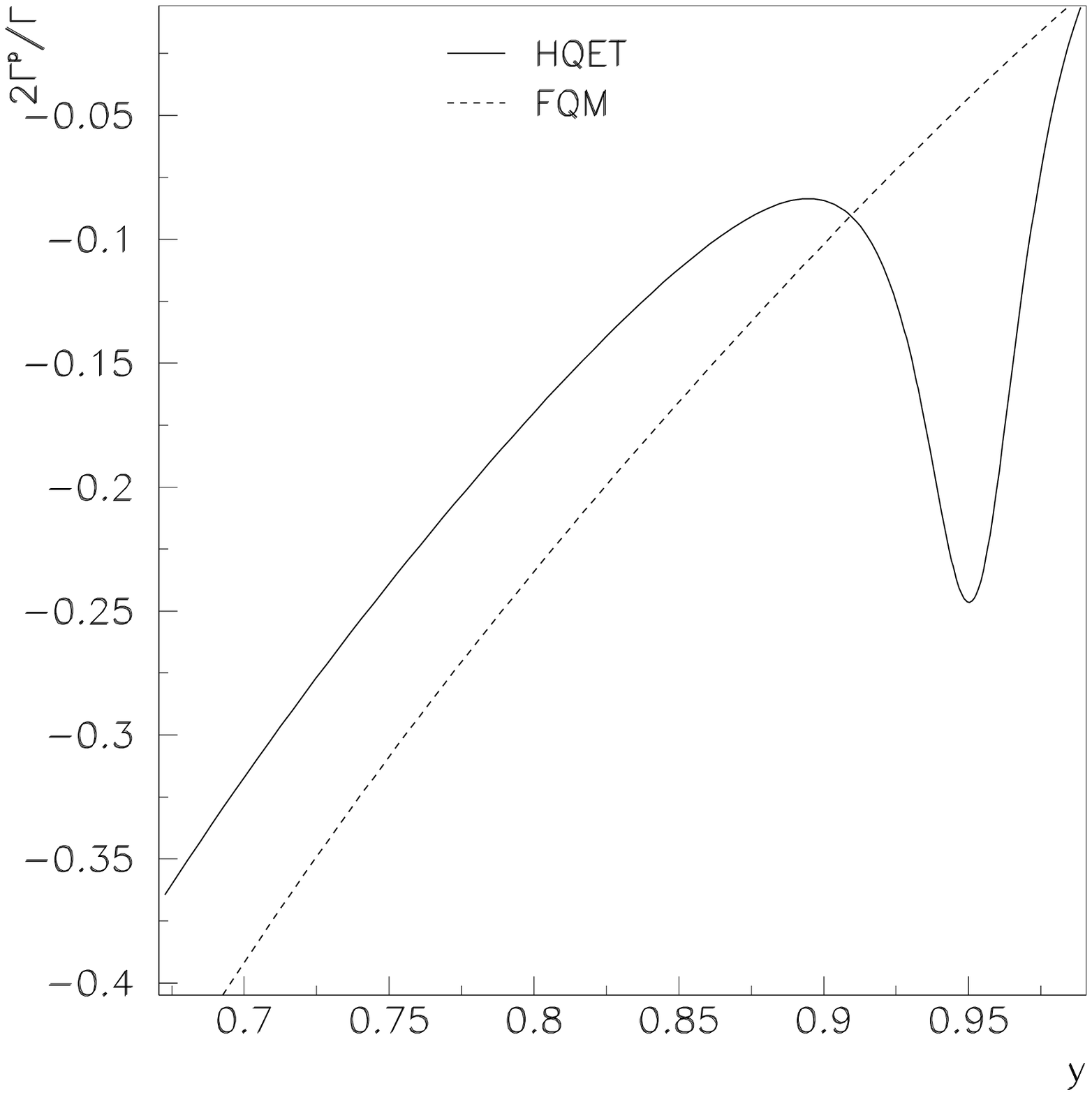}}
\end{center}
\caption{. }
\label{trpl1}
\end{figure}

\begin{figure}[t]
\epsfxsize=352pt
\epsfysize=313pt
\begin{center}
\mbox{\epsffile{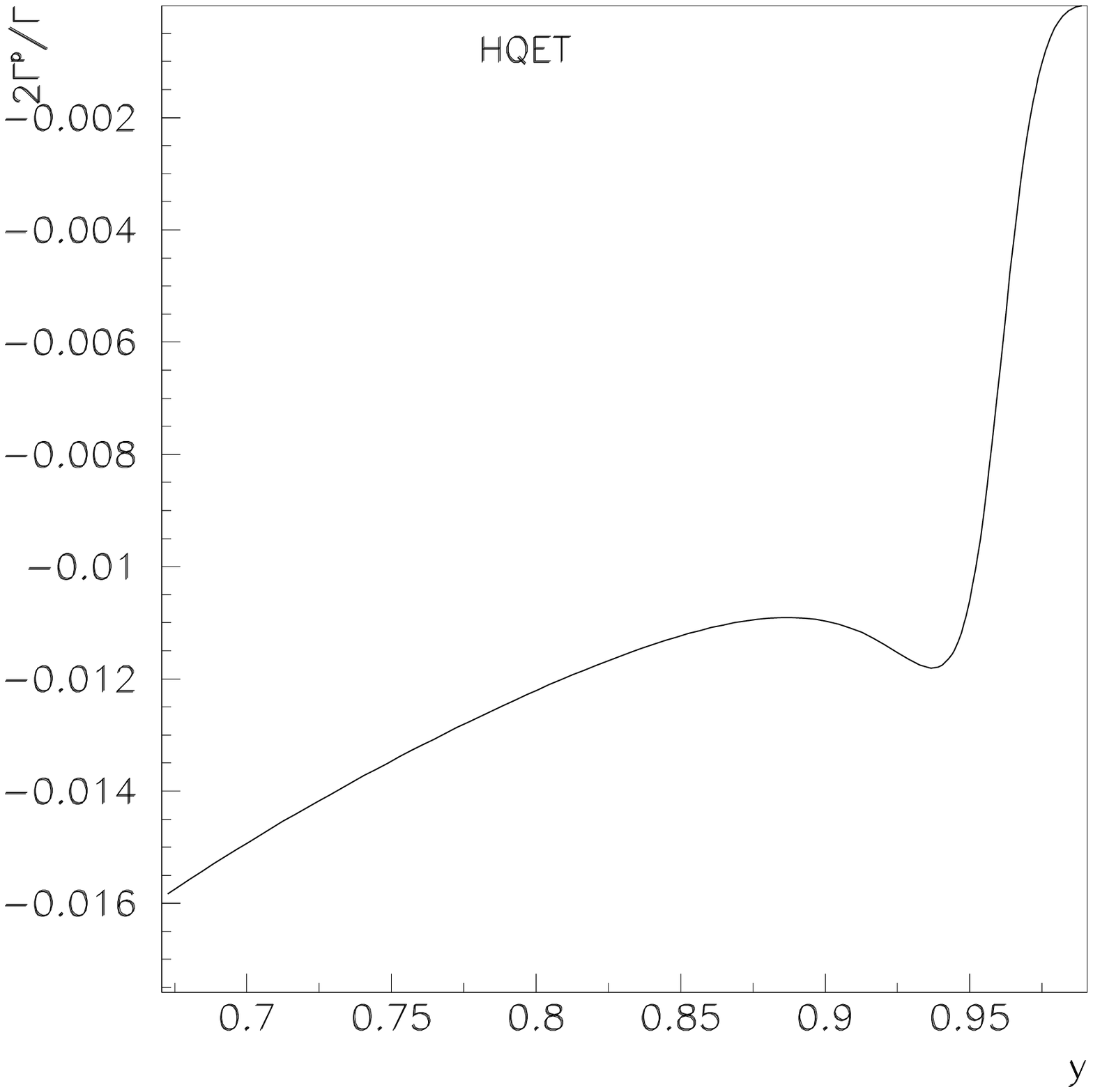}}
\end{center}
\caption{. }
\label{trpl2}
\end{figure}

The transverse polarization in $\Lambda_b$ decays in shown for two different
configurations. Fig. \ref{trpl1} shows it for the hadron spin
parallel to the
$\tau$ spin projection vector and both lying in the decay plane.
The HQET corrections to the FQM
prediction are small. The dip at high $\tau$-energies is again due
to the use of unsmoothed energy spectra. Using smoothed spectra will smear it
out.
For low $\tau$-energies this polarization
is not a small quantity (30\%). As expected, it is of the order $\sqrt{\rt}$.
The
average polarization for this orientation of the spins is
\beq
{\cal P}=-0.19
\eeq
If the hadron spin and the spin projection vector are parallel
to each other and perpendicular
to the decay plane, there are no FQM contributions to the transverse
polarization. All of the polarization shown in Fig. \ref{trpl2} is due to
the HQET corrections. Furthermore, for this choice of the angles, the
expressions containing $\epsilon$ vanish, so that the polarization is
proportional to $K\sqrt{\rt}$. For low $\tau$-energies
a polarization of approximately 1.5\% is expected. The average polarization is
\beq
{\cal P}=-0.011
\eeq
It may be recalled that a small non-zero polarization transverse to the decay
plane can also be induced by final state interactions. That effect, however, is
independent of the spin-orientation of the decaying particle.

\begin{figure}[t]
\epsfxsize=352pt
\epsfysize=313pt
\begin{center}
\mbox{\epsffile{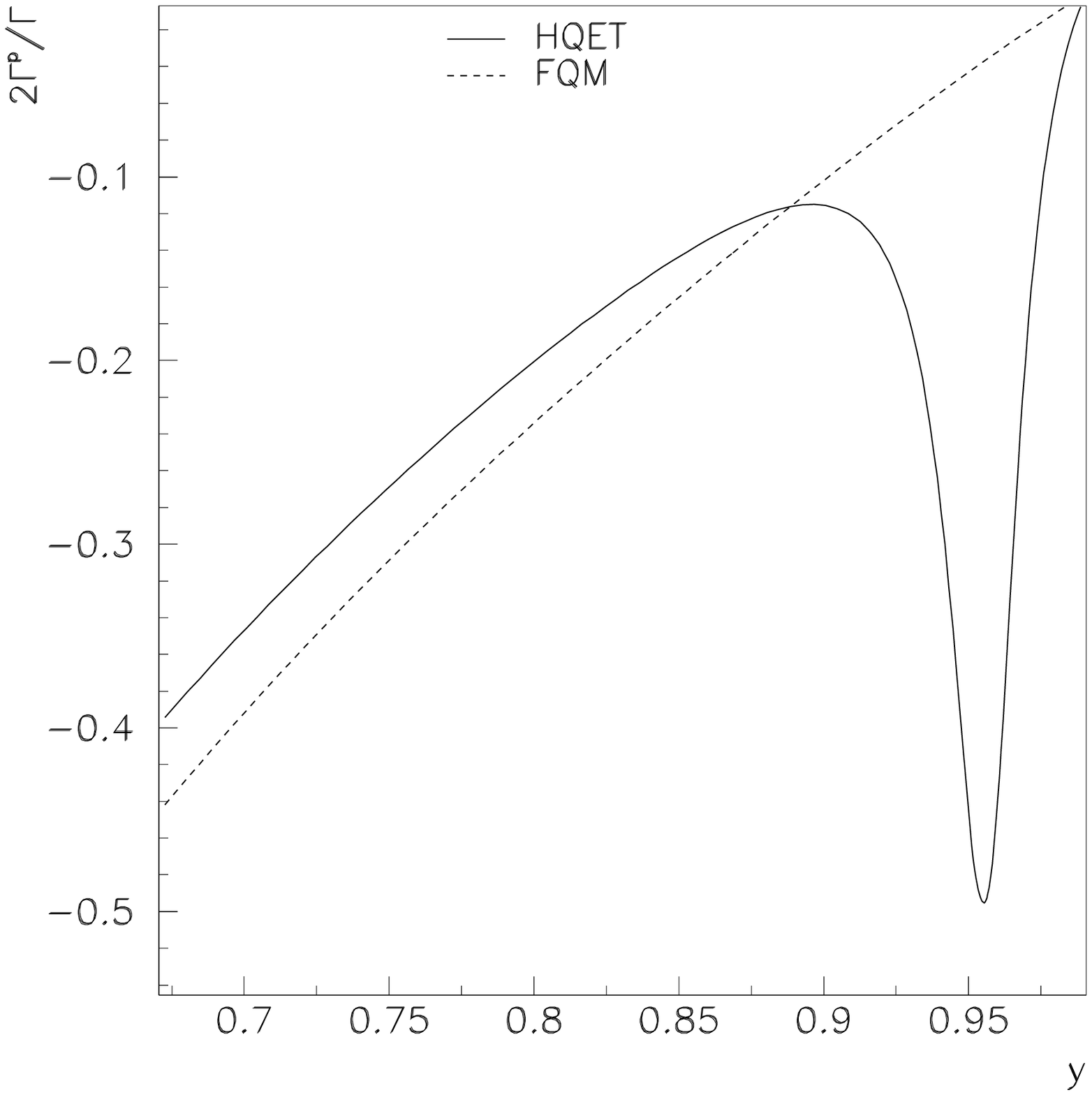}}
\end{center}
\caption{. }
\label{bmpl}
\end{figure}

The transverse polarization in $B$-meson decay is strongest if the spin
projection vector is chosen in the decay plane. Fig. \ref{bmpl} shows
this polarization. In this case the corrections proportional to $G$ have to
be included but parts involving the hadron spin are absent. The prediction
is dominated by the FQM contribution of the order $\sqrt{\rt}$. The dip
at high energies is mostly due to the fact that we use unsmoothed HQET spectra.
The average value for this transverse polarization is
\beq
{\cal P}=-0.2
\eeq
Again, for
low $\tau$-energies, the polarization is rather large (40\%).

\section*{Appendix}

As an example of the fact that no $\de$-functions appear in the two
dimensional decay widths if one parametrizes in terms of $\cos(\te)$ we will
consider the rate for $B \ra X_c e^- \bar{\nu}$ for unpolarized
electrons. It can be decomposed as
\beq
\frac{1}{\Gamma_0}\frac{d\Gamma}{dy\;d\cos(\te)}= D_0+D_1K+D_2G
\eeq
The functions $D_0, D_1$ and $D_2$ are given by
\begin{eqnarray}
D_0&=&48y^2(1-\rj-y)^2\frac{1+\rj+\cos(\te)(1-\rj)}{(2-y+y\cos(\te))^4}\\
D_1&=&\frac{32y^2}{(2-y+y\cos(\te))^6} \left(-2y^3(1-\rj)(1-\rj-y)
	\cos^4(\te) \right. \\
&&\left. \; +y^2((10+2\rj)y^2+(18\rj +3\rj^2-21)y+12(1-rj)^2)\cos^3(\te)
	\right. \nn \\
&&\left. \; y((-10+6\rj)y^3+(21-24\rj+3\rj^2)y^2+12(\rj^2-1)y^2 \right. \nn \\
&&\left. \;\; +4(1+3\rj-9\rj^2+5\rj^3))\cos^2(\te)  \right. \nn \\
&&\left. \; (-(18+10\rj)y^4+(53-22\rj-7\rj^2)y^3-12(5-6\rj+5\rj^2)y^2 \right.
\nn \\
&&\left. \;\;+4(11-16\rj+15\rj^2-10\rj^3)y-16(1-\rj)^2)\cos(\te) \right. \nn \\
&&\left. \; +y(4(4+\rj)y^3-(51-24\rj-3\rj^2)y^2+12(5-4\rj+3\rj^2)y \right. \nn
\\
&&\left. \;\; -4(6-7\rj+6\rj^2-5\rj^3)) \right) \nn \\
D_2&=&\frac{32y^2(1-\rj-y)}{(2-y+y\cos(\te))^5}\left( 2y^2(\rj-2)\cos^3(\te)
	\right.  \\
&&\left.\; +y(2-7y-\rj y -10\rj)\cos^2(\te) \right. \nn \\
&&\left. \; + ((6-4\rj)y^2-4y+12\rj-20\rj^2) \cos(\te) \right. \nn \\
&&\left. \; +((5+3\rj)y^2+(2+10\rj)y-4+8\rj+20\rj^2) \right) \nn
\end{eqnarray}
This decay width is free of $\de$-functions. It can be compared directly
with the corresponding spectrum differential in $y$ and $\qs$ (eqn. 5.2 in
\cite{wise}) which contains $\de$-functions and their first
derivatives. After integration over $\cos(\te)$ and $\qs$ respectively the
results for the lepton energy spectrum $d\Gamma/dy$ coincide.

\newpage

{\bf Figure Captions}

\vspace{0.5cm}

Fig. 1  Branch cuts in the complex x-plane
\vspace{0.3cm}

Fig. 2 Longitudinal Polarization: $\Theta_s=\frac{\pi}{2}$, $\Lambda_b$
	decay
\vspace{0.3cm}

Fig. 3 Transverse Polarization: $\Theta_s=\frac{\pi}{2},
	\de=0,\psi=0$, $\Lambda_b$ decay
\vspace{0.3cm}

Fig. 4 Transverse Polarization: $\Theta_s=\frac{\pi}{2},
	\de=\frac{\pi}{2},\psi=\frac{\pi}{2}$, $\Lambda_b$ decay
\vspace{0.3cm}

Fig. 5 Transverse Polarization: $\de=0$, $B$ decay

\newpage



\newpage


\newpage


\newpage


\newpage


\newpage

\begin{thebibliography}{99}

\bibitem{georgi}
J. Chay, H. Georgi, B.Grinstein, Phys. Lett. B247, 299 (1990)

\bibitem{bigi0}
I.Bigi, M. Shifman, N. Uraltsev, A. Vainshtein, Phys. Rev. Lett. 71, 496 (1993)

\bibitem{wise}
A. V. Manohar and  M.B. Wise, Phys. Rev. D49, 1310 (1994)

\bibitem{blok}
B. Blok, L. Koyrakh, M. Shifman and A.I. Vainshtein, Phys. Rev. D49, 3356
(1994)

\bibitem{kor}
L. Koyrakh, Phys. Rev. D49, 3379 (1994)


\bibitem{korn}
S. Balk, J. G. K\"orner, D. Pirjol, K. Schilcher, Z. Phys. C64, 37 (1994)

\bibitem{falk}
A.F. Falk, Z. Ligeti, M. Neubert, Y. Nir, Phys. Lett. B326, 145 (1994)

\bibitem{lig}
Y. Grossman, Z. Ligeti, hep-ph/9409418

\bibitem{bigi}
I. Bigi, M. Shifman, N.G. Uraltsev, A. Vainshtein, CERN-TH.7250/94
and references therein
\end{thebibliography}
\end{document}